\documentclass[10pt]{article}

\usepackage{amsmath,graphicx}

\def\be{\begin{equation}}
\def\ee{\end{equation}}

\def\e{\text{e}}

\def\d{\text{d}}

\begin{document}
\begin{center}
{\large{\bf Stiff fluid spike solutions from Bianchi type V seed solutions}}
\end{center}

\begin{center}
{\Large\bf }
\vspace{.3in}
{\bf D Gregoris}, 
\\Department of Mathematics and Statistics, Dalhousie University,
\\Halifax, Nova Scotia, Canada B3H 3J5
\\Email:  danielegregoris@libero.it
\vspace{.1in}
\\ {\bf W C Lim},
\\Department of Mathematics and statistics, University of Waikato,
\\ Private Bag 3105, Hamilton 3240, New Zealand
\\Email: woeichet.lim@waikato.ac.nz
\vspace{.1in}
\\ {\bf A A Coley},
\\Department of Mathematics and Statistics, Dalhousie University,
\\Halifax, Nova Scotia, Canada B3H 3J5
\\Email:  aac@mathstat.dal.ca

\vspace{0.2in}

\end{center}

\begin{abstract}
In this paper we expand upon our previous work~\cite{art:Coleyetal2016} by using the entire family of Bianchi type V stiff fluid solutions as
seed solutions of the Stephani transformation. Among the new exact solutions generated, we observe a number of important physical phenomena.
The most interesting phenomenon is exact solutions with intersecting spikes. Other interesting phenomena are solutions with saddle states and a close-to-FL epoch.
\end{abstract}

\section{Introduction}

Spiky structures were first observed in a numerical study by Berger and Moncrief~\cite{art:BergerMoncrief1993}. 
The structures turn out to be permanent spikes, which essentially arise from a discontinuous limit among Kasner solutions towards the initial big bang singularity (at the bang time for brevity).
Further numerical studies reveal recurring transient structures which appear more spiky towards the bang time 
(so-called high-velocity spikes~\cite{art:GarfinkleWeaver2003} and joint low/high velocity spike transitions~\cite{art:Limetal2009,art:HeinzleUgglaLim2012}).
Spiky structures present a possible classical (non-quantum mechanical) way to seed large scale structure formation in the Universe~\cite{art:ColeyLim2012,art:LimColey2014,art:ColeyLim2016}.
The numerical and heuristic analyses in~\cite{art:ColeyLim2012,art:LimColey2014,art:ColeyLim2016} would have been much harder to do without the use of exact solutions.

Indeed, exact solutions are crucial for a better understanding of spiky structures (and for improving the numerical techniques).
The explicit spike solution that encapsulates both transient (high-velocity) and permanent spikes was found by one of the authors~\cite{art:Lim2008}.
This solution was generated by repeatedly applying the Rendall-Weaver transformation~\cite{art:RendallWeaver2001} on the Kasner seed solution.
The Rendall-Weaver transformation is restricted to the family of orthogonally transitive $G_2$ solutions, so other transformations were considered.
The Geroch transformation~\cite{art:Geroch1971,art:Geroch1972} has the advantage that it needs only one Killing vector field for application.
A generalization of the spike solution (which includes joint low/high velocity spike transitions) was generated in~\cite{art:Lim2015} by applying the Geroch transformation on the Kasner seed solution.
The general solution is still a $G_2$ solution, in which spikes form on a plane.

The search for more complex structure led us to apply the 
stiff fluid version of the Geroch transformation (which we shall call the Stephani transformation)~\cite{art:Stephani1988} on various seed solutions.
This includes generating the stiff fluid version of the spike solutions~\cite{art:ColeyLim2016}.
We generated the first $G_1$ spike solution in the recent paper~\cite{art:Coleyetal2016}.
For simplicity we only considered a special case, using a special Bianchi type V seed solution and setting the parameters of the Stephani transformation to special values.
To complete the analysis, we shall now expand the result to the entire family of Bianchi type V seed solutions, and keep all the parameters of the Stephani transformation. 
This generates a family of solutions with a variety of different spiky structures.
The most interesting structure will be the intersecting spikes in the Case I solution with subcritical $p$ and $\omega=0$,
as shown in Figure~\ref{fig:CaseOne_sub_snap}. This solution is the first exact solution with intersecting spikes.

In this paper we will use the signature $(-, +, +, +)$ and geometric units.

\section{The Geroch/Stephani transformation: a review}

In this section we recall the following results by Geroch~\cite{art:Geroch1971,art:Geroch1972} (for vacuum) and Stephani~\cite{art:Stephani1988} (for stiff matter) 
which allow us to generate new exact solutions for the Einstein field equations (EFE) of general relativity.

Let $g_{ab}$ be a known exact solution to the EFE for vacuum or stiff matter (for which the pressure $P$ and the energy density $\rho$ are related by the equation of state $P=\rho$) admitting 
a Killing vector field (KVF) $\xi_a$ orthogonal to the fluid four velocity. Then the steps toward generating another solution ${\tilde g}_{ab}$, which also admits $\xi_a$ as a KVF, are the following.

First, compute the squared norm $\lambda= \xi_a \xi^a$ of the KVF, and integrate the equation $\nabla_a \omega =\epsilon_{abcd} \xi^b \nabla^c \xi^d$
to obtain the twist $\omega$ of the KVF, $\epsilon_{abcd}$ being the completely antisymmetric Levi-Civita symbol.
Next, solve the system of partial derivative equations
\begin{align}
	\nabla_{[a} \alpha_{b]} &= \frac12 \epsilon_{abcd}  \nabla^c \xi^d 
\\
\label{xi_alpha_constraint}
	\xi^a \alpha_a &= \omega
\intertext{for $\alpha_a$, and solve the system}
	\nabla_{[a} \beta_{b]} &= 2\lambda \nabla_a \xi_b +\omega \epsilon_{abcd}  \nabla^c \xi^d 
\\
\label{xi_beta_constraint}
	\xi^a \beta_a &= \omega^2 + \lambda^2 -1
\end{align}
for $\beta_a$.
The squared norm of the KVF in the generated metric, denoted by $\tilde{\lambda}$, is given by
\be
	\tilde{\lambda} = \frac{\lambda}{F},
\ee
where
\be
	F = (\cos\theta- \omega \sin\theta)^2+\lambda^2 \sin^2\theta.
\ee
Construct $\eta_a$ explicitly by
\be
\label{eta_a}
	\eta_a = \frac{\xi_a}{\tilde \lambda}+2 \cos\theta \sin\theta \alpha_a - \sin^2\theta \beta_a.
\ee
Finally, the generated metric is given by explicitly
\be
\label{new1}
{\tilde g}_{ab}= F \left(g_{ab}-\frac{\xi_a \xi_b}{ \lambda} \right)+{\tilde \lambda}\eta_a \eta_b.
\ee
The energy density $\tilde{\rho}$ of the generated solution is given by
\be
	\tilde{\rho}=\frac{\rho}{F}.
\ee

The generated solution is parametrized by the parameter $\theta$. There are also an integration constant $\omega_0$ in $\omega$, and integration ``constant" functions in $\alpha_a$ and $\beta_a$.
The integration ``constant" functions in $\alpha_a$ and $\beta_a$ have no physical effects, and we make the simplest choice possible.
While $\omega_0$ can also be set to zero without loss of generality, we find it simpler to keep $\omega_0$ while setting $\theta=\pi/2$, as we will see below.

\section{The seed solutions considered in this paper}

The family of seed solutions that we shall consider is the Bianchi type V solutions with a non-tilted stiff fluid~\cite{art:EllisMacCallum1969}. 
Its metric and energy density $\rho$ are 
\be
\label{seedG}
 ds^2=\sinh(2t)[-dt^2+dx^2+e^{2x}(\tanh^p(t) dy^2+\tanh^{-p}(t)dz^2)],\quad
        \rho = \frac{3-p^2}{\sinh^3(2t)}.
\ee
From $\rho$ we see that the parameter $p$ takes values between $-\sqrt{3}$ and $\sqrt{3}$. 
The sign of $p$ is not essential. Changing the sign of $p$ switches the role of $y$ and $z$.

There are three special members of this family: $p=0$ gives the isotropic solution (the open Friedmann-Lema\^{\i}tre solution), 
$p=\pm1$ gives the critical anisotropic solution, and $p=\pm\sqrt{3}$ gives the vacuum boundary (the Joseph solution~\cite{art:Joseph1966}).
The first two seed solutions were used in our previous paper~\cite{art:Coleyetal2016}.

The family of seed solutions admit three KVFs, namely:
\be
	\xi_1=\partial_x-y\partial_y-z\partial_z,\quad
	\xi_2=\partial_y,\quad
	\xi_3=\partial_z.
\ee
In order to generate the most general family of solutions, we should use a linear combination of the KVFs
\be
\label{general_KVF}
	\xi = c_1 \xi_1 + c_2 \xi_2 + c_3 \xi_3 = c_1 \partial_x + (c_2-y) \partial_y + (c_3-z) \partial_z.
\ee
Notice that, if $c_1$ is non-zero, then $c_2$ and $c_3$ can be set to zero by a translation in $y$ and $z$, and $c_1$ is set to 1. 
This effectively reduces the linear combination to just $\xi_1$. 
This also means that the case $c_1=0$ needs to be considered separately.
If $c_1$ is zero, then one of $c_2$ and $c_3$ can be set to 1. In the Iwasawa frame that we will adopt later, setting $c_2=1$ results in less artificial frame rotation, so we set $c_2=1$.
So the two KVFs that we shall use are
\begin{align}
\label{KVF1}
\text{Case I:}\quad
 \xi&=\partial_x-y\partial_y-z\partial_z
\\
\label{KVF2}
\text{Case II:}\quad
\xi &= \partial_y + c_3 \partial_z
\end{align}

\section{Full application of the Stephani transformation}

We generate the solutions with the general linear combination of KVFs (\ref{general_KVF}) for the general seed (\ref{seedG}).
As a preliminary computation, we compute $\lambda$ and $\omega$ using the general KVF~(\ref{general_KVF}):
\be
        \lambda = c_1^2 \sinh(2t) + (c_2-c_1 y)^2 \e^{2x} \sinh(2t) \tanh^p t + (c_3-c_1 z)^2 \e^{2x} \sinh(2t) \tanh^{-p} t,
\ee
\be
        \omega = 2p (c_2 - c_1 y) (c_3 - c_1 z) \e^{2x} + \omega_0.
\ee
To complete the transformation, we need to compute $\alpha_a$ and $\beta_a$.
From the constraints (\ref{xi_alpha_constraint}) and (\ref{xi_beta_constraint}) we see that, 
if we perform a change of coordinates to simplify the KVF $\xi_a$ to have only one non-zero component, 
then we would have obtained a component of $\alpha_a$ and $\beta_a$ explicitly.
We shall carry out the change of coordinates before applying the Stephani transformation.

\subsection{Case I: General Bianchi V seed with KVF (\ref{KVF1})}

For Case I, the change of coordinates
\be
\label{coord1}
        (t,x,y,z) = (T,X,Y\e^{-X},Z\e^{-X})
\ee
transforms the KVF (\ref{KVF1}) to
\be
        \xi = \partial_X.
\ee
The metric of the general Bianchi V seed solution (\ref{seedG}) becomes:
\begin{gather}
 g_{00}=-\sinh(2T) \qquad g_{11}= \sinh(2T) \left[ 1 + Y^2 \tanh^p(T) + Z^2 \tanh^{-p}(T) \right] \\
   g_{12}=-Y\sinh(2T)\tanh^p(T)  \qquad g_{13}=-Z \sinh (2T)\tanh^{-p}(T)\\
 g_{22}=\sinh(2T)\tanh^p(T) \qquad g_{33}=\sinh (2T)\tanh^{-p}(T).
\end{gather}
Applying the Stephani transformation, $\lambda$, $\omega$, and the non-zero components of $\alpha_a$ and $\beta_a$ are
\begin{align}
\lambda&= \sinh(2T) \left[ 1 + Y^2 \tanh^p(T) + Z^2 \tanh^{-p}(T) \right]
\\
\omega &= 2pYZ + \omega_0
\end{align}
\be
\alpha_X=\omega, \quad \alpha_Y= Z \cosh(2T), \quad \alpha_Z=-Y \cosh(2T),
\ee
\begin{align}
\beta_X&=\omega^2+\lambda^2-1, \\
\beta_Y&= -Y^3 \sinh^2(2T)\tanh^{2p}(T) -Y\sinh^2(2T)\tanh^p(T)
\notag\\
	&\quad -2pY\int\sinh(2T)\tanh^p(T) \d T 
\notag\\
	&\quad + 2pYZ^2\cosh(2T) - YZ^2\cosh^2(2T) + 2\omega_0 Z \cosh(2T)
\\
 \beta_Z&= -Z^3 \sinh^2(2T)\tanh^{-2p}(T) -Z\sinh^2(2T)\tanh^{-p}(T)
\notag\\
	&\quad +2pZ\int\sinh(2T)\tanh^{-p}(T) \d T 
\notag\\
        &\quad - 2pY^2Z\cosh(2T) - Y^2Z\cosh^2(2T) - 2\omega_0 Y \cosh(2T).
\end{align}
The integrals in $\beta_Y$ and $\beta_Z$ are non-elementary in general. In the special cases $p=0$ (FL seed) and $p=1$ (critical seed) they become elementary.
If $p=0$, the integrals equal $\frac12 \cosh(2T)$. 
If $p=\pm1$, then
\be
	\int\sinh(2T)\tanh^{\pm 1}(T) \d T = \frac12 \sinh(2T) \mp T.
\ee

\subsection{Case II: General Bianchi V seed with KVF (\ref{KVF2})}

For Case II, the change of coordinates
\be
\label{coord2}
        (t,x,y,z) = (T, X, Y, c_3 Y + Z)
\ee
transforms the KVF (\ref{KVF1}) to
\be
        \xi = \partial_Y.
\ee
The metric of the general Bianchi V seed solution (\ref{seedG}) becomes
\begin{gather}
g_{00}=-\sinh(2T) \quad g_{11}=\sinh(2T) \quad g_{23}= c_3 \e^{2X}\sinh(2T)\tanh^{-p}(T) \\
 g_{22}= \e^{2X}\sinh(2T) [ \tanh^p(T)+c_3^2\tanh^{-p}(T) ],\quad
g_{33}=\e^{2X}\sinh(2T) \tanh^{-p}(T).
\end{gather}
Applying the Stephani transformation, $\lambda$, $\omega$, and the non-zero components of $\alpha_a$ and $\beta_a$ are
\begin{align}
\lambda&= \e^{2X}\sinh(2T)[ \tanh^p(T) + c_3^2 \tanh^{-p}(T)]
\\
\omega &= 2 c_3 p \e^{2X} + \omega_0
\end{align}
\be
\alpha_Y= \omega,\quad
\alpha_Z= \e^{2X} [ \cosh(2T) + p ],
\ee
\begin{align}
\beta_Y&= \omega^2 + \lambda^2 -1,
\\
\beta_Z&= c_3 \e^{4X} \sinh^2(2T) [ 1 + c_3^2 \tanh^{-2p}(T) ] +2\e^{2X} [ c_3 p \e^{2X} + \omega_0 ] [ \cosh(2T) + p ].
\end{align}

\subsection{Only one degree of freedom between $\theta$ and $\omega_0$}

From the solutions for $\alpha_a$ and $\beta_a$ above, and from the $\eta_a$ formula~(\ref{eta_a}), 
we see that in the generated metric, $\omega_0$ and $\cos\theta$ always appear together in the expression $(\cos\theta-\omega_0\sin\theta)$.
While $\sin\theta$ can appear alone, it can be absorbed by rescaling the coordinates.
Therefore, there is essentially only one degree of freedom between the two parameters $\theta$ and $\omega_0$.
Two simple choices are $\theta=\pi/2$, which makes $(\cos\theta-\omega_0\sin\theta) = -\omega_0$, and $\omega_0=0$, which makes $(\cos\theta-\omega_0\sin\theta)=\cos\theta$.
The benefits of setting $\theta=\pi/2$ include the decoupling of $\alpha_a$, shorter expressions for solutions, and less demand on computer memory.
We shall set $\theta=\pi/2$ and keep $\omega_0$ free.

For both Case I and Case II solutions above, there are only two essential parameters, namely $\omega_0$ from the transformation and $p$ from the seed solution.

\section{Analysis}\label{sec:analysis}

First we summarize what we did in our previous paper~\cite{art:Coleyetal2016}.
We generated two Case I solutions, one with parameter values $\omega_0=0$ and $p=0$, and the other with parameter values $\omega_0=0$ and $p=1$.
The latter gives a new $G_1$ solution with a permanent spike. We also concluded that the permanent spike forms on a cross. But our analysis was flawed, and we will correct it below.

The expression $F = \omega^2 + \lambda^2$ appears in the denominators of various physical quantities. Its dynamics vary along different worldlines. We now analyze the dynamics of $F$.

\subsection{Early times analysis: Case I}

For Case I,
\be
\label{CaseOne_F}
	F = (2pYZ + \omega_0)^2 + \sinh^2(2T) \left[ 1 + Y^2 \tanh^p(T) + Z^2 \tanh^{-p}(T) \right]^2.
\ee
To study the dynamics of $F$ as $T \rightarrow 0$, we can expand $F$ as a power series in $T$. In particular,
\be
        \sinh^2(2T) \tanh^{\pm 2p}(T) \approx 4 T^{ 2 \pm 2p }.
\ee
We can thus see a few potentially dominant terms -- the $\omega^2$ term, the $4 Y^4 T^{ 2 + 2p }$ or the $4 Z^4 T^{ 2 - 2p }$ term, and the $4T^2$ term.
In the critical $p=\pm1$ case, the $4Z^4$ term and the $4Y^4$ term, respectively, co-dominates with the $\omega$ term.

We see that the parameter $p$ determines which term is dominant. So we split the cases according to its value.

\subsubsection{Supercritical case $|p|>1$}

Along general worldlines, the $4 Z^4 T^{ 2 - 2p }$ term dominates for $p>1$, and the $4 Y^4 T^{ 2 + 2p }$ term dominates for $p<-1$.
Along worldlines where this leading term vanishes ($Z=0$ for $p>1$, and $Y=0$ for $p<-1$), the next dominant term is the $\omega^2$ term, which simplifies to $\omega_0^2$.
In the case $\omega_0=0$, the next dominant term is the $4T^2$ term.

So we see that the spatial profile for $F$ as $T \rightarrow 0$ is discontinuous along $Z=0$ for $p>1$, and along $Y=0$ for $p<-1$.
We conclude that permanent spikes form on these planes.

\subsubsection{Subcritical case $0<|p|<1$}

Along general worldlines, the $\omega^2$ term dominates.
In the case $\omega_0 \neq 0$, the $\omega^2$ term vanishes along two hyperbolic cylinders $2pYZ+\omega_0 =0$, 
where the next dominant term is the $4 Z^4 T^{ 2 - 2p }$ term for $p>0$, and the $4 Y^4 T^{ 2 + 2p }$ term for $p<0$, and these terms do not vanish on the hyperbolic cylinders.
In the case $\omega_0 = 0$, the $\omega^2$ term vanishes along two intersecting planes $Y=0$ and $Z=0$. 
For $p>0$, the next dominant term is $4 Z^4 T^{ 2 - 2p }$ along $Y=0$, $Z\neq0$, and $4T^2$ along $Z=0$.
For $p<0$, the next dominant term is $4 Y^4 T^{ 2 + 2p }$ along $Z=0$, $Y\neq0$, and $4T^2$ along $Y=0$.

So we see that the spatial profile for $F$ as $T \rightarrow 0$ is discontinuous along hyperbolic cylinders $2pYZ+\omega_0 =0$ for the case $\omega_0 \neq 0$,
and along crossing planes $Y=0$ and $Z=0$ for the case $\omega_0 = 0$.
We conclude that permanent spikes form at the hyperbolic cylinders for $\omega_0 \neq 0$, and at the crossing planes for $\omega_0 = 0$.
The intersecting spikes are perhaps the most interesting structure, and will be shown in Figure~\ref{fig:CaseOne_sub_snap}.

\subsubsection{Critical case $p = \pm 1$}

Along general worldlines, the dominant term is $\omega^2 + 4Z^4$ for $p=1$, and $\omega^2 + 4Y^4$ for $p=-1$. 
In the case $\omega_0 \neq 0$, these terms do not vanish.
In the case $\omega_0 = 0$, these terms simplify to $4Z^2(Y^2+Z^2)$ and $4Y^2(Y^2+Z^2)$, respectively, and vanish at $Z=0$ and $Y=0$ respectively, where the next dominant term is $4T^2$.

So we see that the spatial profile for $F$ as $T \rightarrow 0$ is continuous for the case $\omega_0 \neq 0$,
but is discontinuous for the case $\omega_0 = 0$, along $Z=0$ for $p>1$, and along $Y=0$ for $p<-1$.
We conclude that permanent spikes form at these planes for $\omega_0 = 0$.
This corrects the flawed analysis in our previous paper~\cite{art:Coleyetal2016}.

\subsubsection{Special case $p=0$}

Along general worldlines, the dominant term is $\omega_0^2$ if it is non-zero, and $4(1+Y^2+Z^2)T^2$ if $\omega_0=0$.

So we see that the spatial profile for $F$ as $T \rightarrow 0$ is continuous in both cases.
We conclude that no permanent spikes form.

\subsection{Early times analysis: Case II}
        
For Case II,
\be
\label{CaseTwo_F}
        F = (2 c_3 p \e^{2X} + \omega_0)^2 + \e^{4X}\sinh^2(2T)[ \tanh^p(T) + c_3^2\tanh^{-p}(T)]^2.
\ee
The potential dominant terms as $T \rightarrow 0$ are: $(2 c_3 p \e^{2X} + \omega_0)^2$, $4 \e^{4X} T^{2+2p}$ and $4 c_3^4 \e^{4X} T^{2-2p}$.
In the critical $p=\pm1$ case, $4 c_3^2 \e^{4X}$ and $4 \e^{4X}$ respectively co-dominates with $(2 c_3 p \e^{2X} + \omega_0)^2$.

\subsubsection{Supercritical case $|p|>1$}

Along general worldlines, the $4 c_3^4 \e^{4X} T^{ 2 - 2p }$ term dominates for $p>1$, and the $4 \e^{4X} T^{ 2 + 2p }$ term dominates for $p<-1$.
They do not vanish in general. But if $c_3=0$, then for $p>1$ the dominant term is $\omega_0^2$ in general, and $4 \e^{4X} T^{2+2p}$ if $\omega_0=0$.

In all cases, spatial profile for $F$ as $T \rightarrow 0$ is continuous. We conclude that no permanent spikes form.

\subsubsection{Subcritical case $0<|p|<1$}

Along general worldlines, the $\omega^2$ term dominates. The $\omega^2$ term vanishes at the plane $X = \frac12 \ln \left( \frac{-\omega_0}{2 c_3 p} \right)$, if the quantity in the bracket is positive.
Then at the plane, the dominant term is $4 c_3^4 \e^{4X} T^{2-2p}$ for $p>0$, and $4 \e^{4X} T^{2+2p}$ for $p<0$. 
They do not vanish in general. If $c_3=0$ the $\omega^2$ term becomes $\omega_0^2$. And if $\omega_0$ is also zero, then the dominant term is $4 \e^{4X} T^{2+2p}$.

We see that the spatial profile for $F$ as $T \rightarrow 0$ is discontinuous at the plane $X = \frac12 \ln \left( \frac{-\omega_0}{2 c_3 p} \right)$, if the quantity in the bracket is positive.
We conclude that permanent spikes form in this typical (but not general) case.

\subsubsection{Critical case $p = \pm 1$}

Along general worldlines, for $p=1$ the dominant term is $(2 c_3 \e^{2X} + \omega_0)^2 + 4 c_3^4 \e^{4X}$, which vanishes if $c_3=0=\omega_0$. Then the dominant term is $4 \e^{4X} T^4$.
For $p=-1$ the dominant term is $(-2 c_3 \e^{2X} + \omega_0)^2 + 4 \e^{4X}$, which does not vanish.

In all cases, the spatial profile for $F$ as $T \rightarrow 0$ is continuous. We conclude that no permanent spikes form.

\subsubsection{Special case $p = 0$}

Along general worldlines, the dominant term is $\omega_0^2$. If $\omega_0=0$ then the dominant term is $ 4 \e^{4X} [ 1 + c_3^2 ]^2 T^2$.

In both cases, the spatial profile for $F$ as $T \rightarrow 0$ is continuous. We conclude that no permanent spikes form.

\subsection{Late times analysis}

As $T \rightarrow \infty$, the dominant term is $[ 1 + Y^2 + Z^2 ]^2 \sinh^2(2T)$ for Case I, and
$\e^{4X} [ 1 + c_3^2 ]^2 \sinh^2(2T)$ for Case II.
The spatial profile for $F$ as $T \rightarrow \infty$ is continuous. We conclude that no permanent spikes form.

\subsection{Intermediate times analysis}

We do not have a good strategy to proceed with an intermediate times analysis. In the next section, we will present numerical plots to help guide our analysis.

\section{Numerical evidence}

Even if we could analyze the intermediate times dynamics of $F$, the dynamics of the generated solution is more complex than just the dynamics of one quantity.
For example, the Hubble expansion $H$ is given by:
\be
	H = \frac{G}{3\sinh^{3/2}(2T) F^{3/2}},
\ee
where $F$ is given by (\ref{CaseOne_F}) and (\ref{CaseTwo_F}), respectively, for Case I and Case II, and $G$ is given by
\begin{align}
        G &= 3\cosh(2T) \omega^2 + 5\cosh(2T) \lambda^2
\notag\\
        &\quad + 2p\sinh(2T) [Y^2 \tanh^p(T) - Z^2 \tanh^{-p}(T)] \lambda
\end{align}
for Case I, and
\begin{align}
	G &= 3\cosh(2T) \omega^2 + 5\cosh(2T) \lambda^2
\notag\\
        &\quad + 2p\e^{2X}\sinh(2T) [\tanh^p(T) - c_3^2\tanh^{-p}(T)] \lambda
\end{align}
for Case II. Then the density parameter $\Omega$ is
\be
	\Omega = \frac{\rho}{3H^2} = 3(3-p^2) \frac{F^2}{G^2}.
\ee
Notice that Hubble-normalized physical quantities now have $G$ rather than $F$ in the denominator.
$1/G$ has possibly different spiky structures than $1/F$. As a result, Hubble-normalized physical quantities have possibly different structures than unnormalized ones.
$\Omega$, or the ratio $F/G$, captures the difference in structures of $F$ and $G$.

Before we carry out more analysis, it is a good idea to produce numerical plots of important quantities to better understand the dynamics of the solution. 

\begin{figure}
\begin{center}
\includegraphics[width=\linewidth]{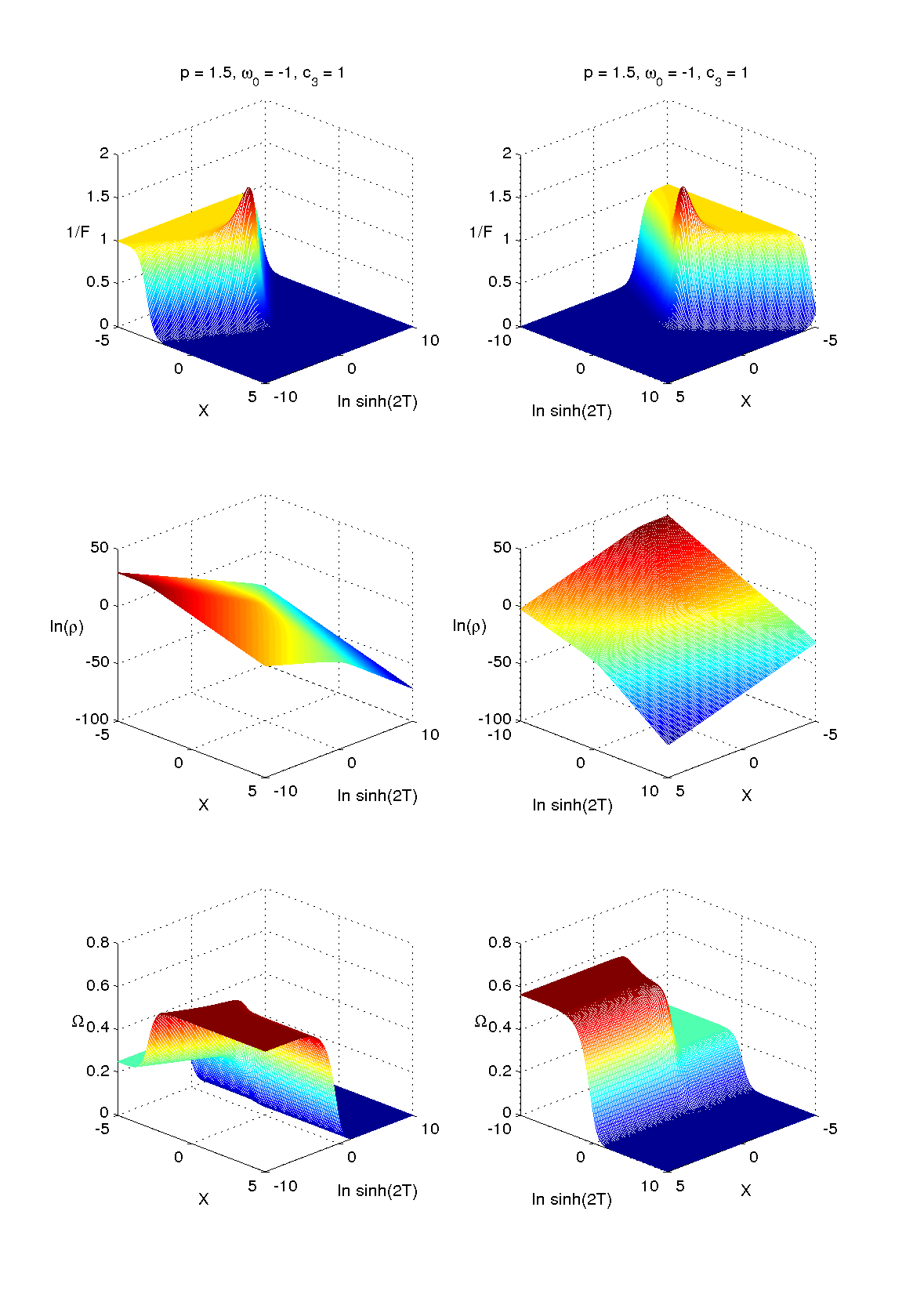}
\end{center}
\caption{Case II solution with supercritical $p$ and $p \omega_0 c_3 <0$. 
Notice the global maximum of $1/F$ at intermediate time, as well as the different dynamics of $\Omega$ along different worldlines.}
\label{fig:CaseTwo_super}
\end{figure}

\begin{figure}
\begin{center}
\includegraphics[width=\linewidth]{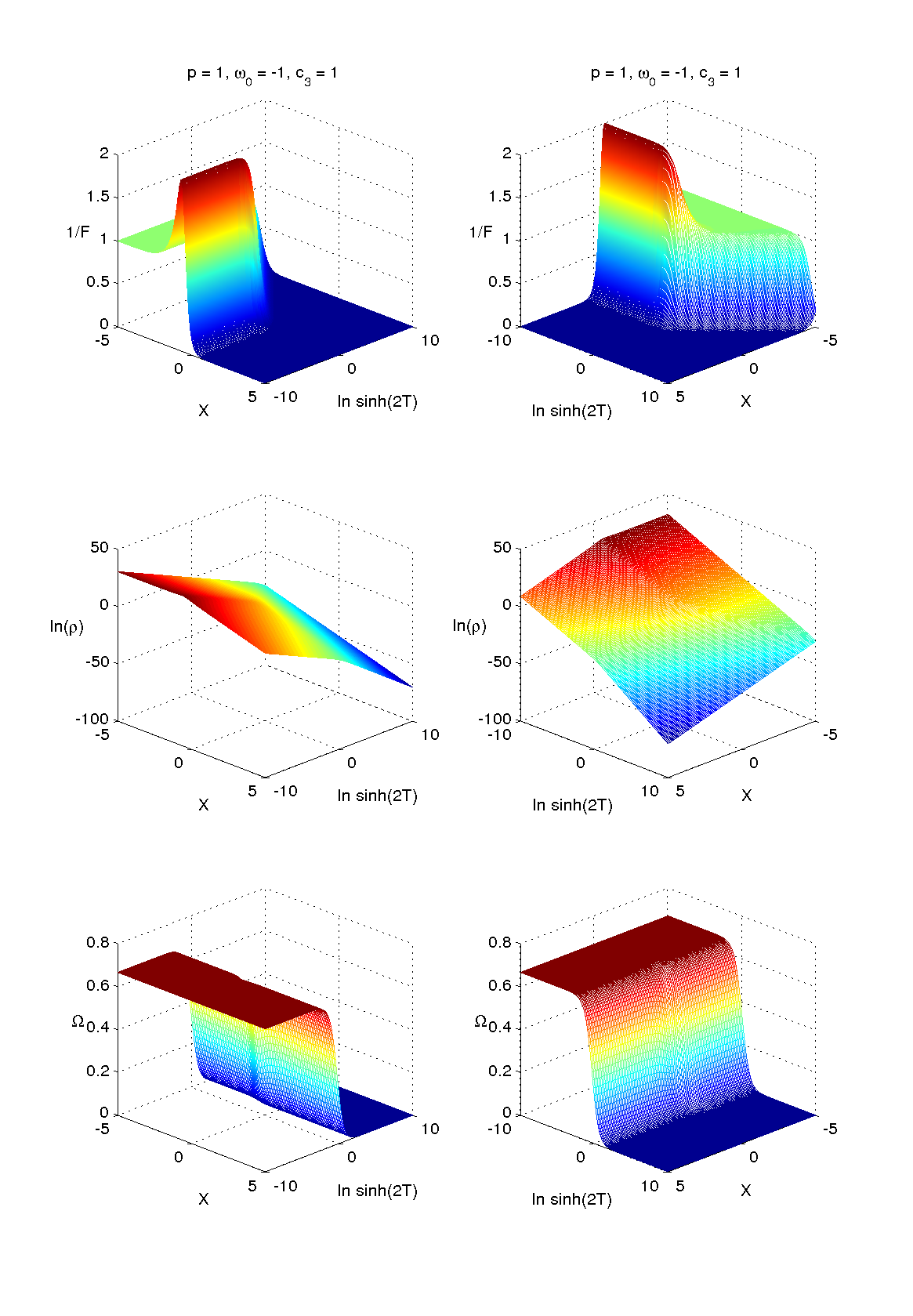}
\end{center}
\caption{Case II solution with critical $p$ and $p \omega_0 c_3 <0$. Notice the global maximum of $1/F$ at the bang time.}
\label{fig:CaseTwo_crit}
\end{figure}

\begin{figure}
\begin{center}
\includegraphics[width=\linewidth]{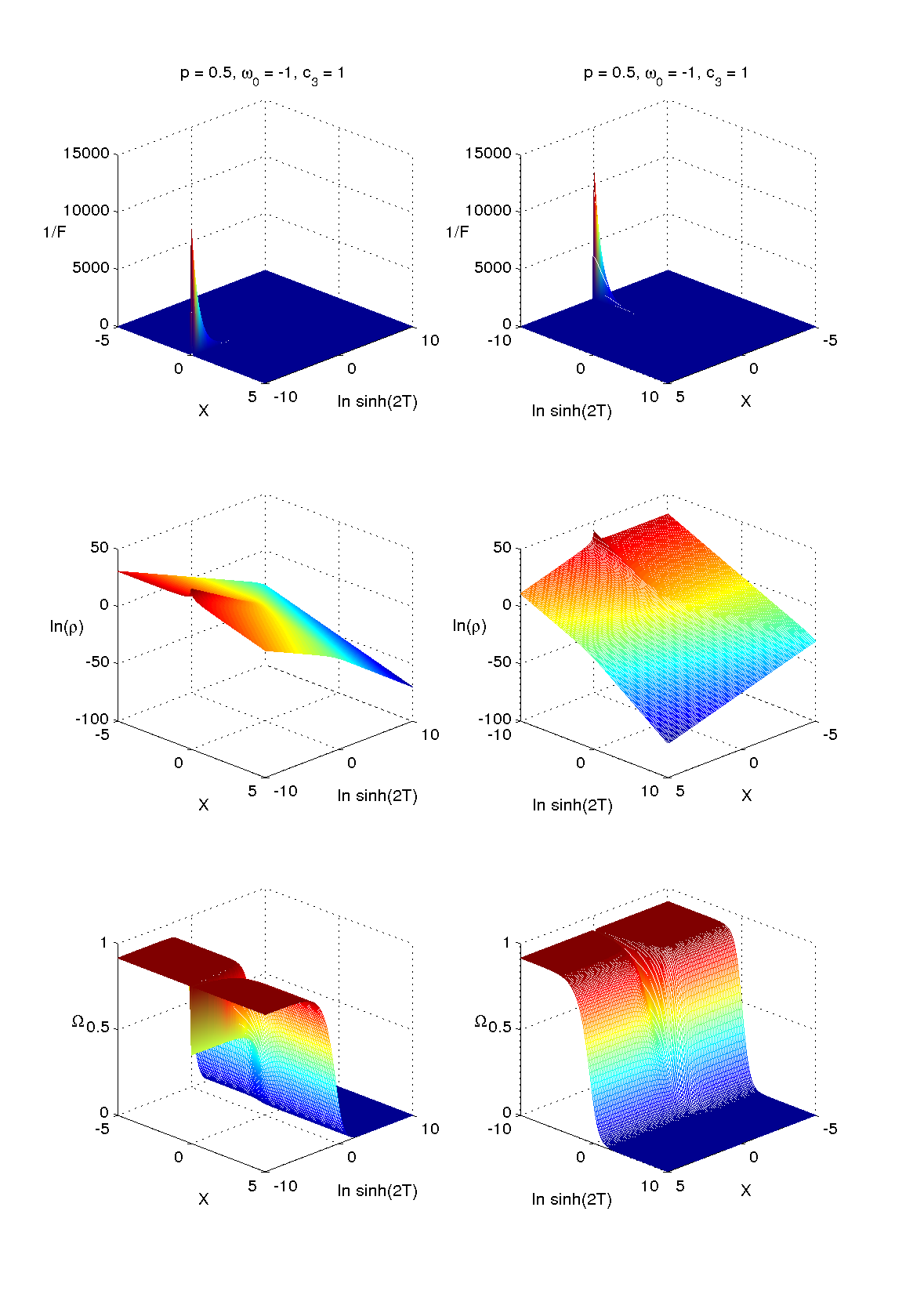}
\end{center}
\caption{Case II solution with subcritical $p$ and $p \omega_0 c_3 <0$.}
\label{fig:CaseTwo_sub}
\end{figure}

\begin{figure}
\begin{center}
\includegraphics[width=\linewidth]{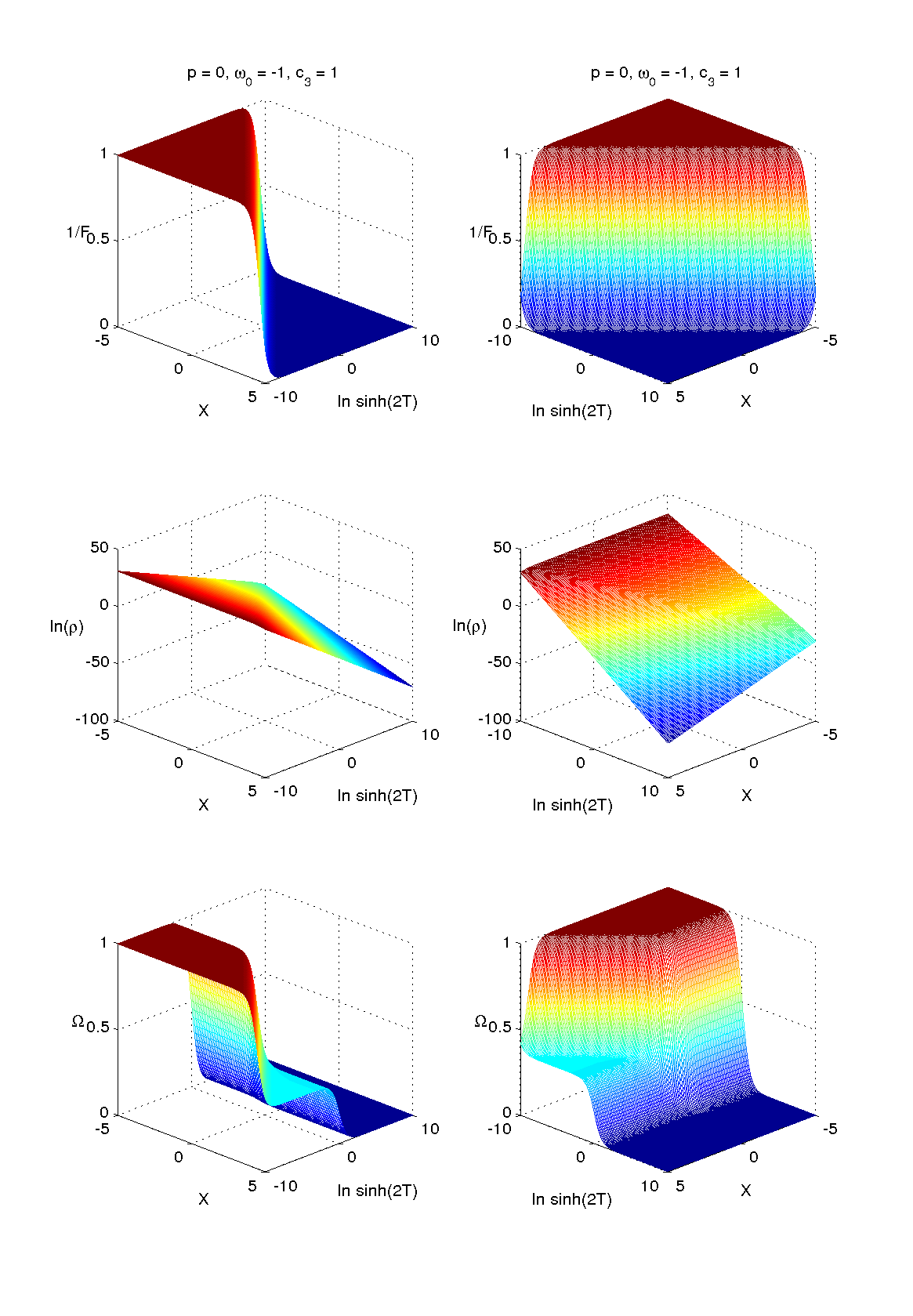}
\end{center}
\caption{Case II solution with $p=0$ and $\omega_0\neq0$. Notice the different dynamics of $\Omega$ along different worldlines.}
\label{fig:CaseTwo_p0}
\end{figure}

Case II solutions are the simpler ones, as they depend on two variables, $T$ and $X$. We can plot the graph of a quantity against $\ln \sinh (2T)$ and $X$.
In Figures~\ref{fig:CaseTwo_super}--\ref{fig:CaseTwo_p0}, we plot $1/F$, $\ln \rho$ and $\Omega$. Views from two different angles are plotted.
The choice of parameter values is guided by the cases discussed in the previous section. 
Rather than showing an exhaustive list of cases, we show only cases that have different representative behaviours.

\begin{figure}
\begin{center}
\includegraphics[width=\linewidth]{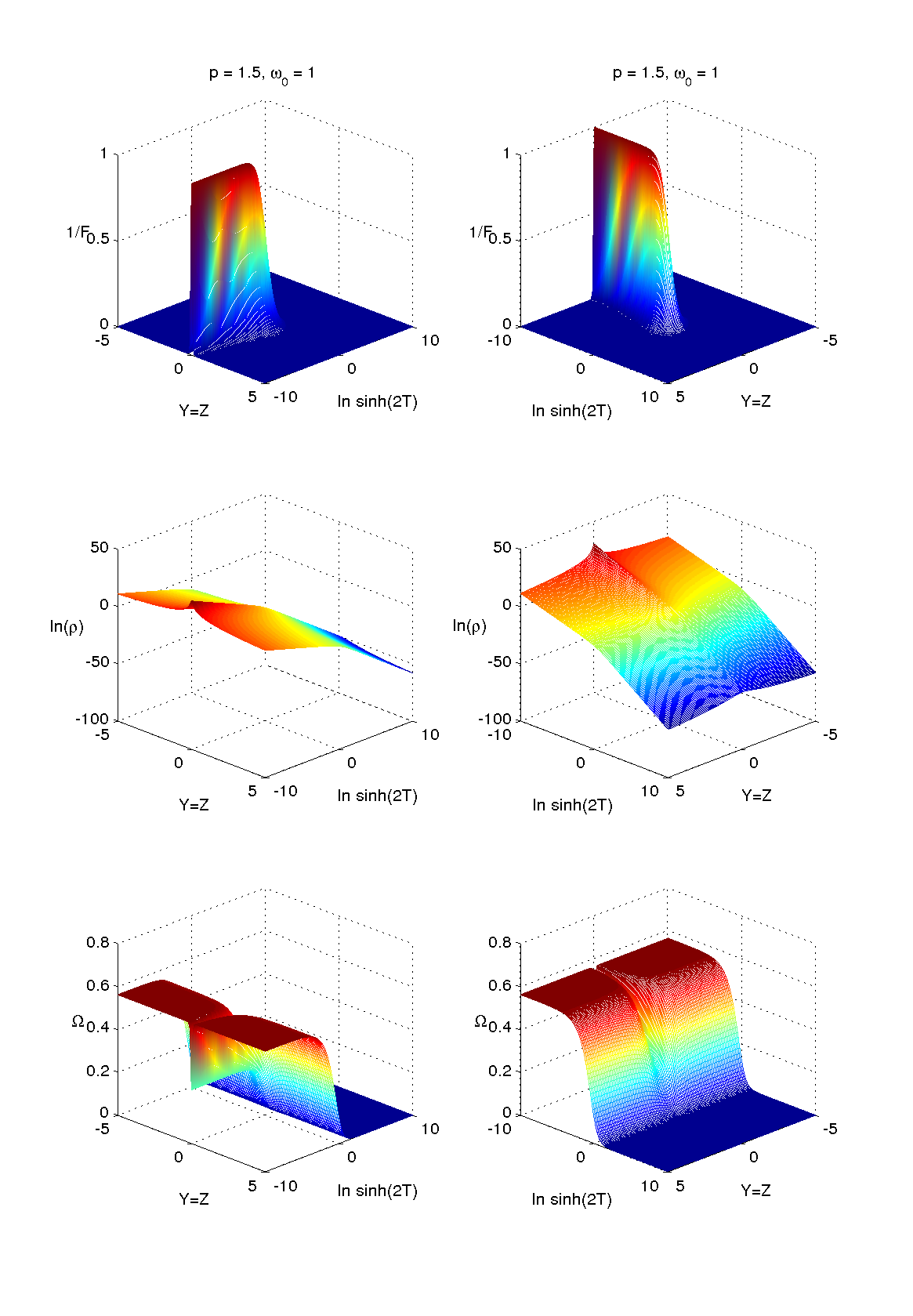}
\end{center}
\caption{Case I solution with supercritical $p$. A permanent spike forms.}
\label{fig:CaseOne_super}
\end{figure}

\begin{figure}
\begin{center}
\includegraphics[width=\linewidth]{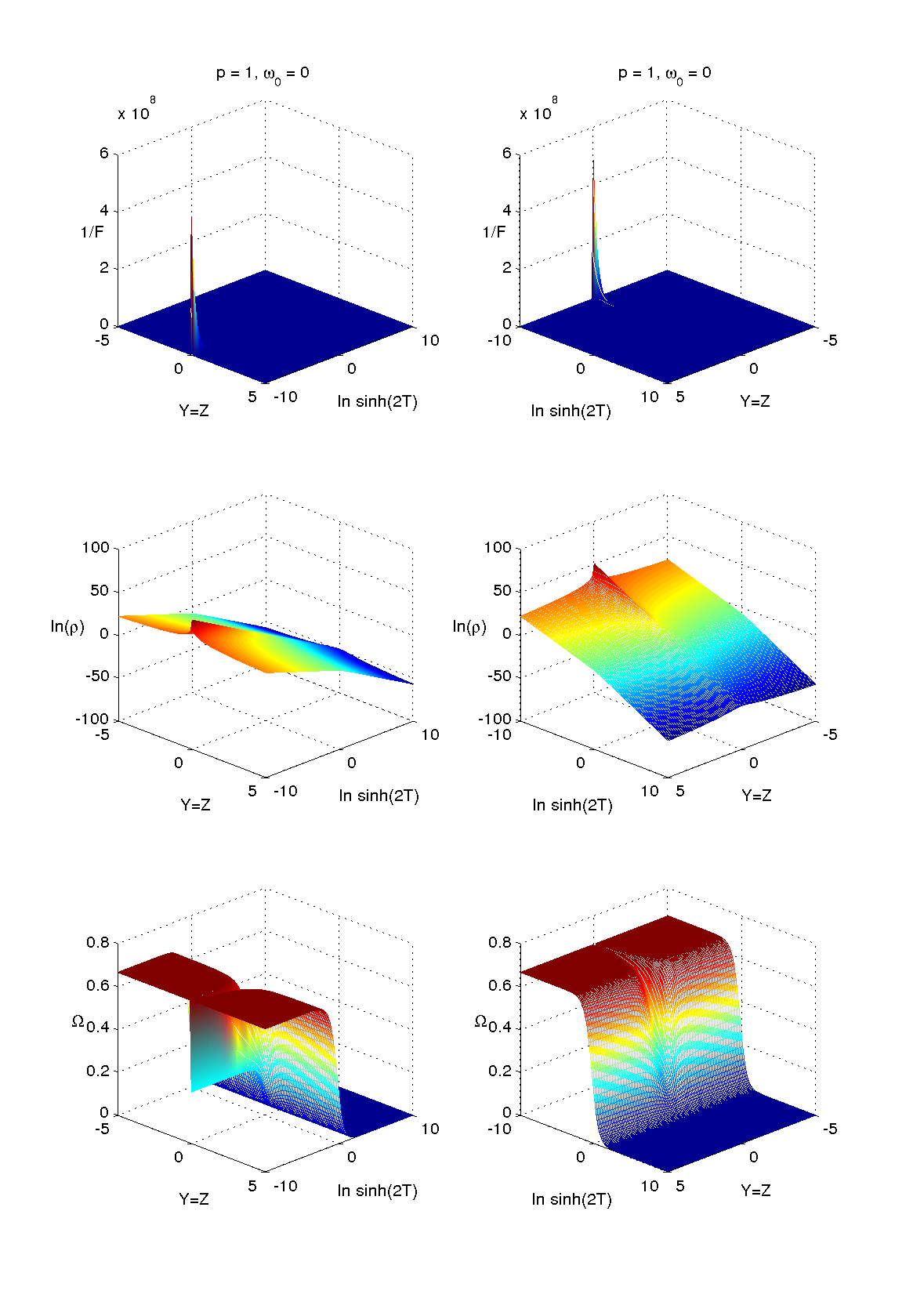}
\end{center}
\caption{Case I solution with critical $p$, $\omega_0=0$.}
\label{fig:CaseOne_crit}
\end{figure}

\begin{figure}
\begin{center}
\includegraphics[width=\linewidth]{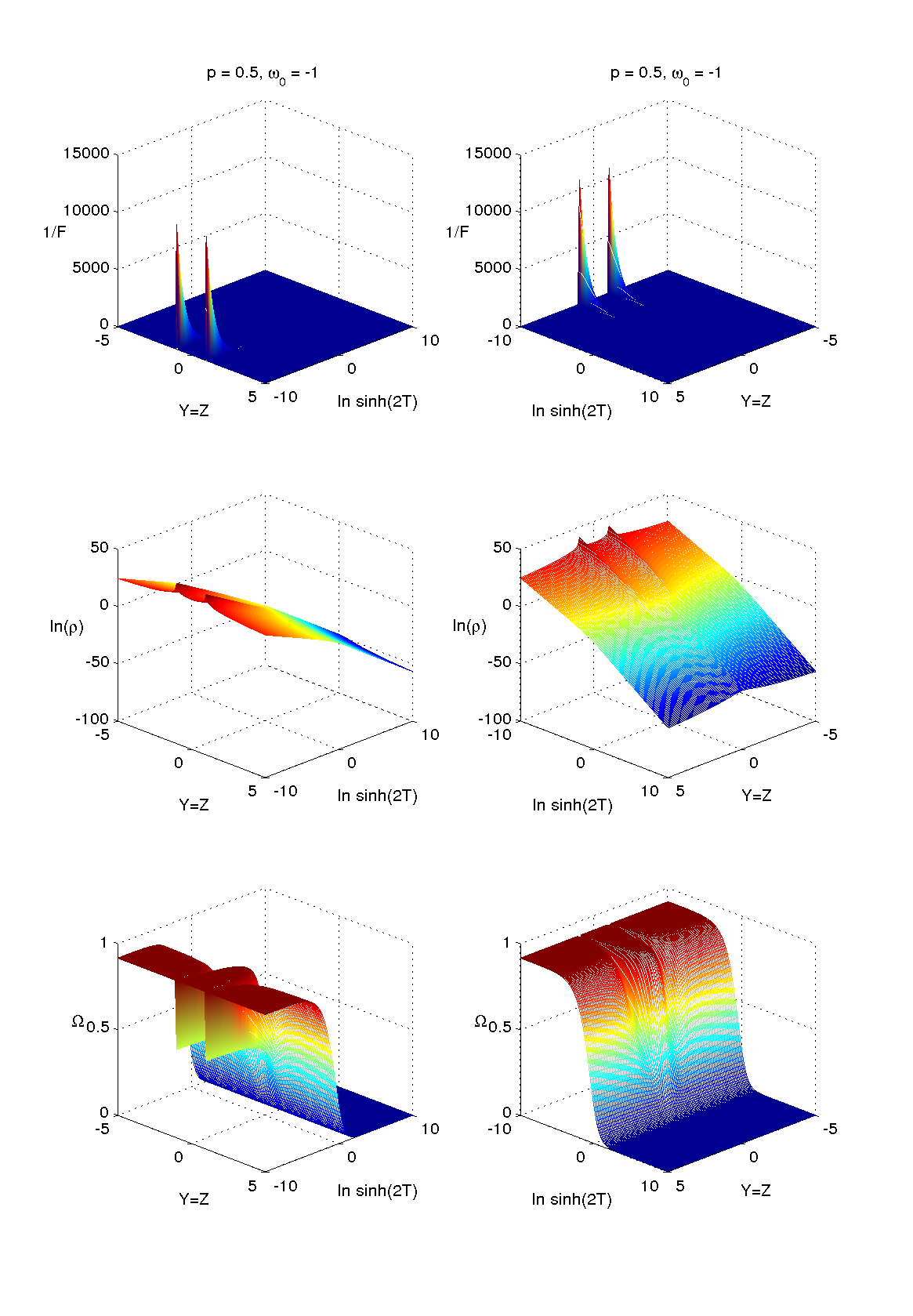}
\end{center}
\caption{Case I solution with subcritical $p$, $\omega_0\neq0$. Two spikes.}
\label{fig:CaseOne_sub}
\end{figure}

\begin{figure}
\begin{center}
\includegraphics[width=\linewidth]{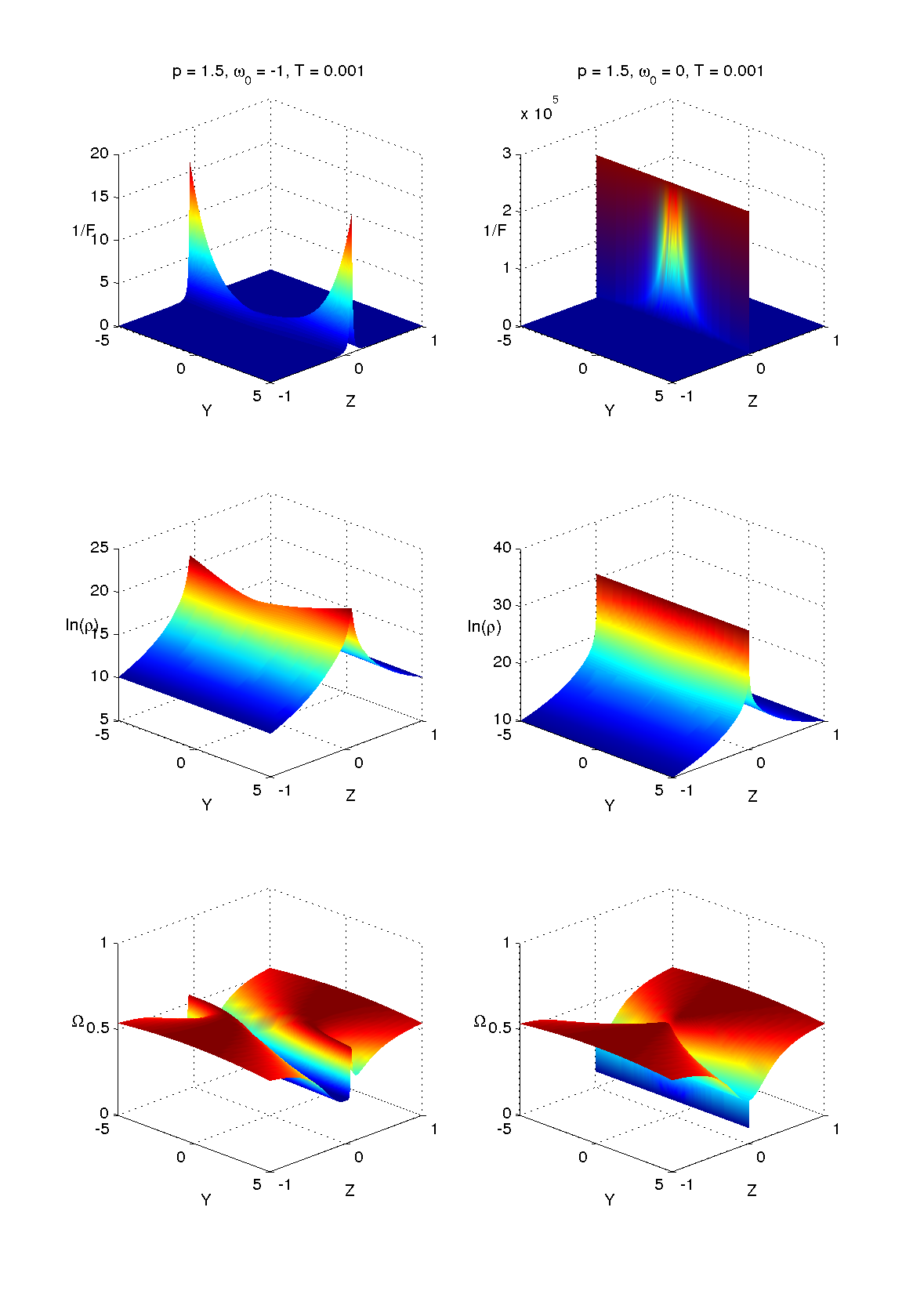}
\end{center}
\caption{Case I solution with supercritical $p$. Snapshot shows spikes and interesting structures in $\Omega$.}
\label{fig:CaseOne_super_snap}
\end{figure}

\begin{figure}
\begin{center}
\includegraphics[width=\linewidth]{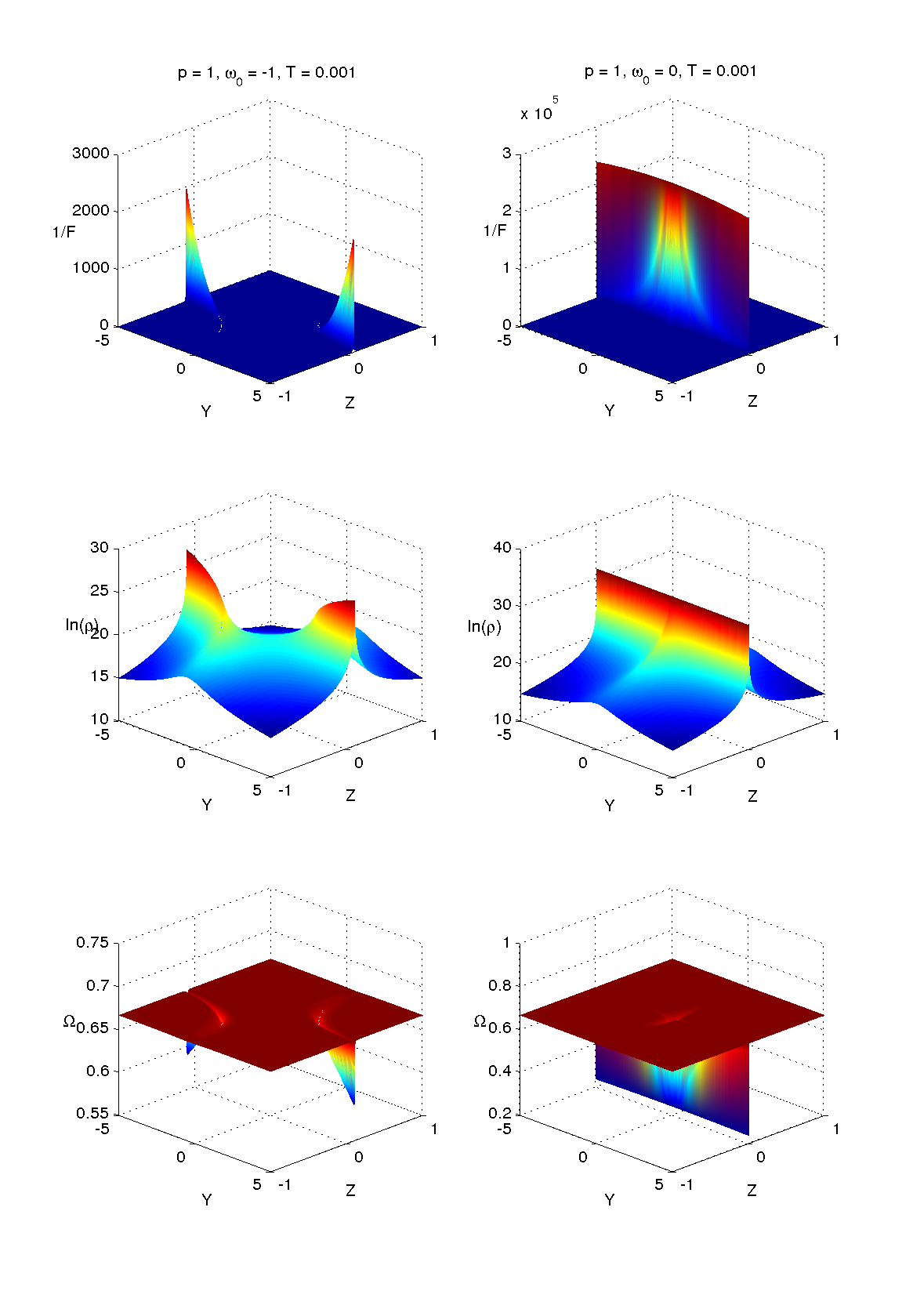}
\end{center}
\caption{Case I solution with critical $p$. Snapshot shows spikes in $\Omega$.}
\label{fig:CaseOne_crit_snap}
\end{figure}

\begin{figure}
\begin{center}
\includegraphics[width=\linewidth]{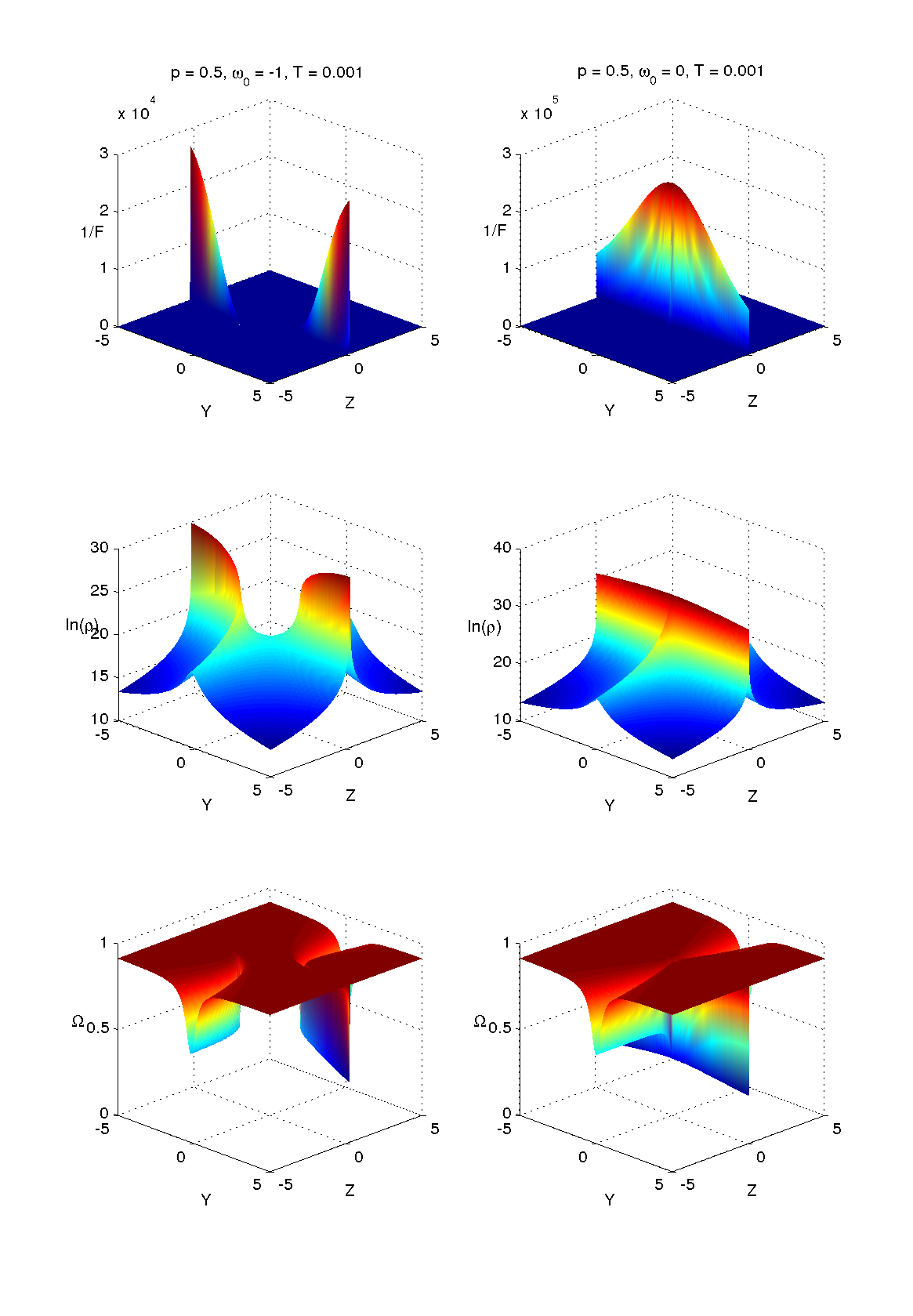}
\end{center}
\caption{Case I solution with subcritical $p$. Snapshot shows two spikes (non-intersecting for $\omega_0\neq0$ and intersecting for $\omega_0=0$).}
\label{fig:CaseOne_sub_snap}
\end{figure}

Case I solutions depend on $T$, $Y$ and $Z$. We can suppress one of the variables and plot the graph of a quantity against the other two variables.
In Figures~\ref{fig:CaseOne_super}--\ref{fig:CaseOne_sub}, we plot $1/F$, $\ln \rho$ and $\Omega$ againts $\ln \sinh(2T)$ and $Y=Z$. Views from two different angles are plotted.
Then in Figures~\ref{fig:CaseOne_super_snap}--\ref{fig:CaseOne_sub_snap}, we plot snapshots of $1/F$, $\ln \rho$ and $\Omega$ againts $Y$ and $Z$ at fixed $T$ close to the bang time.
Special care is taken to improve the numerical resolution around the narrow spikes in these snapshots.

Let us now discuss the figures, in turn.
Figure~\ref{fig:CaseTwo_super} represents the Case II solution with supercritical $p$ and $p \omega_0 c_3 <0$.
Two interesting features are the global maximum of $1/F$ at intermediate time, and the different dynamics of $\Omega$ along different worldlines.
For $X<0$, there is an additional saddle state for $\Omega$.

Figure~\ref{fig:CaseTwo_crit} represents the Case II solution with critical $p$ and $p \omega_0 c_3 <0$.
The global maximum of $1/F$ is now located at the bang time.

Figure~\ref{fig:CaseTwo_sub} represents the Case II solution with subcritical $p$ and $p \omega_0 c_3 <0$. A permanent spike forms at early times.

Figure~\ref{fig:CaseTwo_p0} represents the Case II solution with $p=0$ and $\omega_0 \neq 0$. 
This solution is past asymptotic to the flat Friedmann-Lema\^{\i}tre (FL) solution.
Notice the different dynamics of $\Omega$ along different worldlines.
For $X>0$, there is an additional saddle state for $\Omega$.

Moving on to the Case I solutions, Figure~\ref{fig:CaseOne_super} represents the Case I solution with supercritical $p$. 
In the view with $Y=Z$, a permanent spike forms.
The snapshot in Figure~\ref{fig:CaseOne_super_snap} reveals richer spatial structure. In addition to the narrow spike, there are interesting milder inhomogeneous structures.

Figure~\ref{fig:CaseOne_crit} represents the Case I solution with critical $p$ and $\omega_0=0$. A permanent spike forms.
The snapshot in Figure~\ref{fig:CaseOne_crit_snap} reveals different spike geometry for $\omega_0\neq0$ and for $\omega_0=0$.
The latter case was studied in our previous paper~\cite{art:Coleyetal2016}, and our flawed analysis there erroneously concluded that the spikes form a cross.
Spike crossing actually occurs only in the case discussed next.

Figure~\ref{fig:CaseOne_sub} represents the Case I solution with subcritical $p$.
In the view with $Y=Z$, two permanent spike forms.
The snapshot in Figure~\ref{fig:CaseOne_sub_snap} reveals richer spatial structure. 
The spikes form on hyperbolas for the case $\omega_0\neq0$, and on intersecting planes $YZ=0$ for the case $\omega_0=0$.
The intersecting spikes are the most interesting phenomenon that we observe in the generated solutions.
The Case I solution with subcritical $p$ and $\omega_0=0$ is the first exact solution with intersecting spikes.
Observe that the energy density $\rho$ is higher on the spikes, and highest at the intersection.
Thus the intersecting spikes exemplify prototypical intersecting walls that divide the space into four regions with a lower density (prototypical voids).
Upon further analysis, we find that the limit of $\Omega$ as $T\rightarrow0$, for $0<p<1$ and $\omega_0=0$, is
\be
	\Omega \rightarrow \begin{cases}
				\dfrac{3}{25}(3-p^2) & \text{along $Z=0$}
				\\
				\dfrac{3(3-p^2)}{(5-2p)^2} & \text{along $Y=0$, $Z\neq0$}
				\\
				\dfrac{3-p^2}{3} & \text{along $Y\neq0$, $Z\neq0$.}
				\end{cases}
\ee

\section{Close-to-FL solutions}

There are other physical quantities for which we have not presented numerical plots. Among them are the shear components and the spatial curvature.
A set of interesting solutions are those that are close to FL (close to isotropic). 
Figure~\ref{fig:CaseTwo_p0} shows that $\Omega$ tends to 1 at the bang time, indicating that the solution is past asymptotic to the flat FL solution.
Subcritical Case I and Case II solutions with $p \ll 1$, $\omega_0 \neq 0$ have $\Omega$ that tends to a limit close to 1 at early times.

\begin{figure}
\begin{center}
\includegraphics[width=\linewidth]{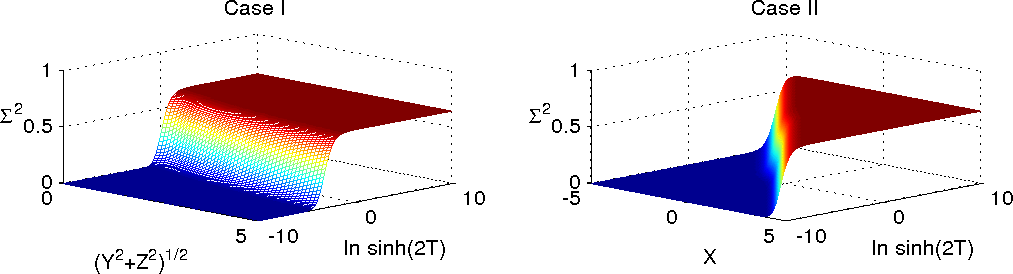}
\end{center}
\caption{Hubble-normalized shear scalar $\Sigma^2$ for Case I and Case II ($c_3=1$), with $p=0$ and $\omega_0=1$.}
\label{fig:shear}
\end{figure}

Let us discuss this in more detail. For the $p=0$ solutions, the expressions for Hubble-normalized shear components are manageable. 
Using the Iwasawa frame (frame rotation $R_1 = -\Sigma_{23}$, $R_2=\Sigma_{13}$, $R_3=-\Sigma_{12}$)~\cite{art:HeinzleUgglaLim2012}, the shear components 
for Case I $p=0$ solutions are
\begin{align}
	 -\frac12 \Sigma_{11}  &= \Sigma_{22} = \Sigma_{33} = \frac{4\lambda^2}{3\omega_0^2 + 5\lambda^2}
\\
	\Sigma_{12} &= -6 \omega_0 \sinh(2T)\tanh(2T) \frac{Z(1+Y^2+Z^2)}{\sqrt{1+Z^2}(3\omega_0^2 + 5\lambda^2)}
\\
	\Sigma_{13} &=  6 \omega_0 \sinh(2T)\tanh(2T) \frac{Y\sqrt{1+Y^2+Z^2}}{\sqrt{1+Z^2}(3\omega_0^2 + 5\lambda^2)}
\\
	\Sigma_{23} &= 0,
\end{align}
where $\lambda = \sinh(2T) (1+Y^2+Z^2)$.
The shear scalar $\Sigma^2 = \frac16 \Sigma_{ab} \Sigma^{ab}$ is then
\be
	\Sigma^2 = \frac{4\lambda}{(3\omega_0^2+5\lambda^2)^2} \left[ 4\lambda^3 + 3\omega_0^2\tanh^2(2T)\sinh(2T)(Y^2+Z^2)\right].
\ee
For Case II, we have $\lambda = (1+c_3^2) \e^{2X} \sinh(2T)$,
\begin{align}
	-\frac12 \Sigma_{22} &= \Sigma_{11} = \Sigma_{33} = \frac{4\lambda^2}{3\omega_0^2 + 5\lambda^2}
\\
	\Sigma_{12} &= \Sigma_{13} = 0
\\
	\Sigma_{23} &= - \frac{6 \omega_0 \tanh(2T) \lambda}{3\omega_0^2 + 5\lambda^2}
\\
	\Sigma^2 &=  \frac{4\lambda^2 [3\omega_0^2\tanh^2(2T) + 4\lambda^2]}{(3\omega_0^2+5\lambda^2)^2}.
\end{align}
Figure~\ref{fig:shear} plots the shear scalar for each case with non-zero $\omega_0$. 
For Case I, $\Sigma^2$ quickly transitions from zero to $\frac{16}{25}$ (anisotropic at late times).
But for Case II, worldlines with negative $X$ experience a delay in such transition.
So Case II provides a more interesting example in this respect.

\section{Conclusion}

In this paper we have expanded our previous work~\cite{art:Coleyetal2016} using the entire family of Bianchi type V solutions as
seed solutions of the Stephani transformation. Among the newly generated solutions, we observed a number of interesting phenomena.

The most interesting phenomenon is the intersecting spikes in the Case I solution with subcritical $p$ and $\omega=0$,
as shown in Figure~\ref{fig:CaseOne_sub_snap}.
This solution is the first exact solution with intersecting spikes. In this case the intersecting planes are $Y=0$ and $Z=0$.
Intersecting spikes exemplify a prototypical intersecting walls, with the density higher on the walls, and highest at the intersection. 
The intersecting walls divide the space into four regions with a lower density (prototypical voids).
We note that the Universe is dominated by bubbles of large voids surrounded by denser walls~\cite{art:ColeyWiltshire2017}.
It has also been argued that some of the large scale observational anomalies may require the existence of non-linear structures early in the Universe~\cite{art:dolgov2016}.
For example, most galactic nuclei contain supermassive black holes which must have already existed by a redshift of $\sim 10$.
However, such enormous black holes could not have naturally formed so early within the conventional cosmological model,
and there must have already existed large seed black holes well before galaxy formation (with these seeds subsequently growing through accretion).
In~\cite{art:LimColey2014} it was suggested that intersecting spikes may present an interesting model for the formation of such non-linear structures at late times.
However, we note that in the new exact solutions presented here there are no permanent spikes at late times. 
For future works we hope to generate solutions with spikes at late times.

Another interesting phenomenon is the existence of saddle states for certain worldlines, as seen in
Figures~\ref{fig:CaseTwo_super} and~\ref{fig:CaseTwo_p0}.
A third interesting phenomenon is the close-to-FL epoch at early times in $p=0$ solutions, as seen in
Figures~\ref{fig:CaseTwo_p0} and~\ref{fig:shear}. In addition, there are worldlines with arbitrarily long close-to-FL epoch.
This may be of interest for researchers working on perturbation theory who require an exact solution to compare against.
Most interesting perhaps, Case I solution with subcritical $p\approx0$ and $\omega=0$
shows two phenomena at the same time -- it has a spike crossing with a close-to-FL background at early times.
For future works as regards close-to-FL epoch, we hope to generate solutions that are close to FL at intermediate times or late times.
Solutions close to FL at intermediate times might be suitable models of the Universe.
Solutions close to FL at late times are of interest regarding the question of late-time isotropization.

The outlook is promising. As mentioned in the discussion section of our previous paper~\cite{art:Coleyetal2016},
there are many possible seed solutions that one can try (for examples, Bianchi type VI and VII solutions, Szekeres solutions, and even the solutions generated here). 
There may yet be more surprises about spikes and spike crossings.

\section*{Acknowledgments}

This work was supported, in part, by NSERC of Canada and AARMS. The symbolic computation software {\tt MAPLE} is essential in carrying out the computation.
The numerical software {\tt MATLAB} is used to plot the figures.

\end{document}